\documentclass[aps,pra,twocolumn,superscriptaddress]{revtex4-1}

\usepackage{hyperref}
\hypersetup{
     colorlinks   = true,
     citecolor    = red}
\usepackage{amsmath}
\usepackage{bm}
\usepackage{gensymb}
\usepackage[normalem]{ulem}
\usepackage{euscript}
\usepackage[pdftex]{graphicx}
\usepackage{xspace}
\usepackage{xfrac}
\usepackage{enumerate}
\usepackage{braket}
\usepackage{float,fancyvrb}
\usepackage[T1]{fontenc}
\usepackage{lmodern}
\graphicspath{{./Figures/}}
\usepackage{wrapfig}
\usepackage{scrextend}
\usepackage{comment}
\usepackage{url}
\usepackage{amssymb}
\usepackage{xcolor} 
\usepackage{lineno} 

\newcommand{\uu}[1]{\ensuremath{\, \mathrm{#1}}} 

\definecolor{ao}{rgb}{0.0, 0.5, 0.0}

\begin{document}

\title{Floquet-engineered quantum state manipulation in a noisy qubit}

\author{Eric Boyers}
\affiliation{Department of Physics, Boston University, Boston, Massachusetts 02215, USA}
\affiliation{these authors contributed equally to this work}
\author{Mohit Pandey}
\affiliation{Department of Physics, Boston University, Boston, Massachusetts 02215, USA}
\affiliation{these authors contributed equally to this work}
\author{David K. Campbell}
\affiliation{Department of Physics, Boston University, Boston, Massachusetts 02215, USA}
\author{Anatoli Polkovnikov}
\affiliation{Department of Physics, Boston University, Boston, Massachusetts 02215, USA}
\author{Dries Sels}
\affiliation{Department of Physics, Harvard University, Cambridge, MA 02138, USA}
\affiliation{Theory of quantum and complex systems, Universiteit Antwerpen, B-2610 Antwerpen, Belgium}
\author{Alexander O. Sushkov}
\email{asu@bu.edu}
\affiliation{Department of Physics, Boston University, Boston, Massachusetts 02215, USA}
\affiliation{Department of Electrical and Computer Engineering, Boston University, Boston, Massachusetts 02215, USA}
\affiliation{Photonics Center, Boston University, Boston, Massachusetts 02215, USA}

\date{\today}

\begin{abstract}
Adiabatic evolution is a common strategy for manipulating quantum states and has been employed in diverse fields such as quantum simulation, computation and annealing. 
However, adiabatic evolution is inherently slow and therefore susceptible to decoherence. 
Existing methods for speeding up adiabatic evolution require complex many-body operators or are difficult to construct for multi-level systems. 
Using the tools of Floquet engineering, 
we design a scheme for high-fidelity quantum state manipulation, utilizing only the interactions available in the original Hamiltonian. We apply this approach to a qubit and experimentally demonstrate its performance with the electronic spin of a Nitrogen-vacancy center in diamond. Our Floquet-engineered protocol achieves state preparation fidelity of $0.994 \pm 0.004$, on the same level as the conventional fast-forward protocol, but is more robust to external noise acting on the qubit. Floquet engineering provides a powerful platform for high-fidelity quantum state manipulation in complex and noisy quantum systems.
\end{abstract}

\maketitle
\hypersetup{linkcolor=green}
\section{Introduction}
Accurate manipulation of quantum systems is a fundamental goal in many areas of quantum science, ranging from quantum information science through  quantum simulation to quantum sensing. Control over the quantum state of a system is crucial as a preparatory step for a subsequent computation or simulation~\cite{bloch2008}, or as a goal in itself, as in adiabatic quantum computation~\cite{farhi2000quantum,albash218}.
Some quantum states are ``easy'' to prepare, for example, by cooling the system to the ground state of its Hamiltonian. However, a number of applications require access to quantum states that are ``difficult'' to prepare with high fidelity. For example, quantum annealing ~\cite{santoro2006optimization,santoro2008optimization} requires finding the quantum ground state of a complex many-body Hamiltonian, and entanglement-assisted quantum sensing requires preparation of entangled states of large numbers of qubits in order to achieve sensitivity beyond the standard quantum limit~\cite{Jones2009}.
One of the standard approaches to state preparation is to use adiabatic evolution: initialize the system in an eigenstate of a simple, easy to prepare Hamiltonian and then adiabatically change the Hamiltonian to a new one with one of the eigenstates (typically the ground state) being the desired target state. This approach has been used for transport of ultra-cold atoms~\cite{krinner2014}, many-body state engineering~\cite{Bernien2017}, and quantum thermodynamics ~\cite{salamon2009maximum,rezek2009quantum}. 

Adiabatic evolution is a generic strategy, but the evolution rate must be much smaller than the energy gaps in the system. Therefore, this approach is slow and susceptible to decoherence due to inevitable interactions with the environment~\cite{Huang2011}. Shortcuts to adiabaticity are methods of achieving faster adiabatic evolution, in order to maintain high fidelity in the presence of decoherence and noise. One technique is counter-diabatic (also known as transitionless) driving~\cite{berry2009transitionless, demirplak2003adiabatic,demirplak2005assisted,Bason2012,Zhang2013}. In this approach, transitionless evolution at arbitrary velocities is achieved by adding additional velocity-dependent counter terms to the Hamiltonian. However, for complex quantum  many-body systems, these additional counter terms  are, in general, highly non-local operators and, as a result, typically experimentally inaccessible. A related technique involves fast-forward driving  protocols, which use only operators available in the original Hamiltonian but employ more complex time dependence to achieve high fidelity~\cite{masuda2009fast}.
Two main strategies have been proposed for finding fast-forward protocols. The first uses methods from optimal control theory to analytically or numerically find driving protocols that achieve near-unit fidelity~\cite{Judson1992,Caneva2011,glaser, hegerfeldt, farhi}. Although successful in many cases, such protocols are hard to compute for generic quantum systems~\cite{day}. The second strategy is based on a recently-proven statement that any fast-forward drive can be obtained as a unitary transformation of a counter-diabatic drive \cite{bukov2018geometric,petiziol2018fast}. In this approach, the problem of finding a fast-forward protocol can be decomposed into finding a counter-diabatic protocol first, and then finding the time-dependent unitary transformation that converts the counter-diabatic Hamiltonian into the original one, with modified time-dependent couplings. However, once again, there is no general method for finding this transformation for many-body systems.

\begin{figure}[ht]
	\centering
	\includegraphics[width= 0.5\textwidth]{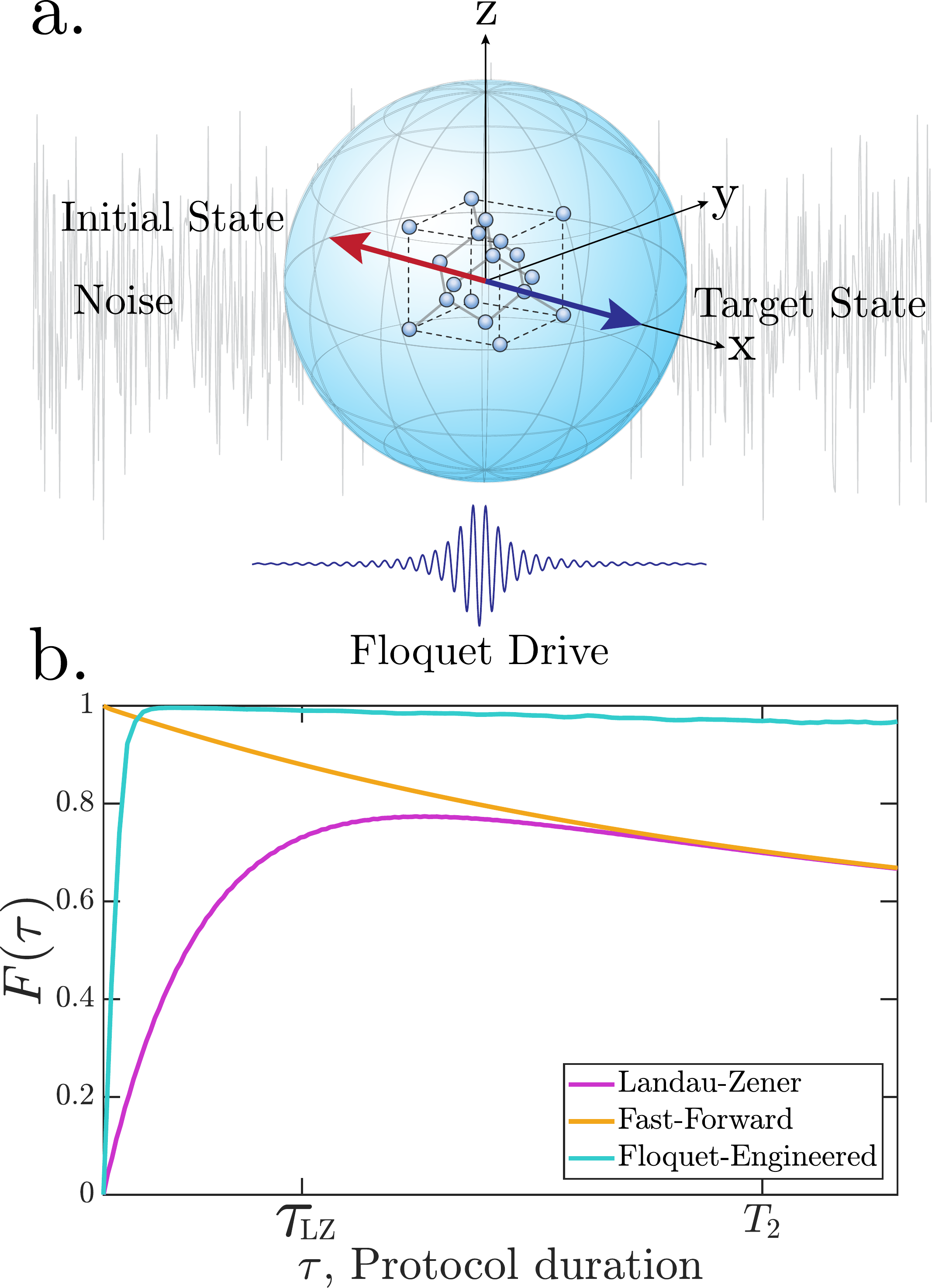}
	\caption{(a) Schematic for quantum state manipulation of a qubit implemented with the electronic spin of a nitrogen-vacancy center in diamond. Our protocol includes a high-frequency Floquet drive that allows faster transitionless evolution, and maintains robust performance in presence of external noise. (b) Numerical calculation of the fidelity for preparing final state $|\psi_f\rangle$ for varying protocol duration, in presence of decoherence. A linear Landau-Zener ramp (purple) has poor fidelity at short times due to transitions and at long times due to decoherence. A fast-forward drive (orange) removes Landau-Zener transitions, but is still susceptible to decoherence at long protocol durations. A Floquet-engineered fast-forward drive (blue) suppresses Landau-Zener transitions and decouples the system from decoherence, resulting in high fidelity over a broad range of protocol durations. Ticks on the x-axis indicate the minimum time to achieve adiabatic evolution according to the Landau-Zener condition: $\tau_{\text{LZ}}=\lambda/\Delta^2$, and the spin coherence time $T_2$.}
	\label{fig:fig1}
\end{figure}

Here we use Floquet engineering to construct an approximate fast-forward protocol from a counter-diabatic protocol.
By driving the system at high frequency we introduce a time-scale separation between the periodic modulation and the change of the protocol control fields. This separation allows us to construct the aforementioned transformation and results in protocols that become asymptotically exact in the limit of infinite frequency. This Floquet-engineered fast-forward driving can achieve nearly unit fidelity with a target state for short protocol duration and protects the quantum system against decoherence, as illustrated in fig.~\ref{fig:fig1}. In this work we apply this approach to a qubit, but the methodology can be generalized to more complex quantum systems by including higher harmonics of the fundamental Floquet frequency (see appendix \ref{append.derivation}).  To demonstrate the feasibility of the suggested approach, we experimentally implemented the Floquet-engineered protocol in a single qubit based on a nitrogen-vacancy (NV) center in diamond and compared its performance with the conventional fast-forward protocol in presence of external noise.

\section{Model}
Consider a two-level quantum system, or qubit, with the following Hamiltonian
\begin{equation} \label{eq:01}
H(t)= \Delta \sigma_z + \lambda (t)\sigma_x
\end{equation}
where the $\sigma_{x,y,z}$ are Pauli matrices, and we work in units with $\hbar=1$ throughout the paper, i.e. energies are measured in Hz. For a spin-1/2 system, $\Delta$ is proportional to the magnitude of the static magnetic field in the $z$-direction, and $\lambda (t)$ is proportional to the magnitude of the time-dependent magnetic field in $x$-direction, serving as the external control parameter. The initial state $|\psi_0\rangle=|-x\rangle$ and the target state $|\psi_t\rangle=|x\rangle$ are eigenstates of $\sigma_x$, see fig.~\ref{fig:fig1}(a). The  fidelity of the protocol is defined as the overlap of the final spin state $|\psi(t)\rangle$ with the target state: $F(t) = |\langle\psi(t)|\psi_t\rangle|^2$. The initial and target states are adiabatically connected but separated by an avoided crossing at $\lambda=0$. 
For an adiabatic protocol the relative change of the instantaneous gap has to be much smaller than the gap: $\dot{\lambda}/\Delta\ll\Delta$. This puts a strong constraint on the minimal required time to implement an adiabatic linear sweep protocol: $ \tau \gg \lambda/\Delta^2$. If this time is comparable to, or longer than the decoherence time of the qubit, the adiabatic protocol never achieves high fidelity, as in fig.~\ref{fig:fig1} (b). 

A counter-diabatic protocol introduces an additional control field that keeps the system in the instantaneous ground state~\cite{berry2009transitionless, kolodrubetz2017geometry}:

\begin{align}
H_\mathrm{CD}(t) =  \Delta\sigma_z + \lambda (t)  \sigma_x+  \dfrac{1}{2} \dfrac{\Delta \dot{\lambda}}{\Delta^2+ \lambda^2} \sigma_y.
\label{CD_eqn}
\end{align}
For a qubit, the $\sigma_y$ control, corresponding to a time-dependent magnetic field in $y$-direction for a spin-1/2, is as easy to implement as $\sigma_x$. However, for generic many-body systems the counter-diabatic Hamiltonian would require access to a large number of multi-qubit operators, that, in general, are experimentally inaccessible. Fast-forward protocols avoid this complication~\cite{bukov2018geometric,masuda2009fast} by performing a virtual rotation around the x-axis, producing a control Hamiltonian that involves only the control fields 
$B_z, B_x$ corresponding to the original operators $\sigma_z,\sigma_x$:

\begin{align} \label{eq:FF}
H_{\text{FF}} &= B_z(t)\sigma_z + B_x(t)\sigma_x  \\
&= \Delta \sqrt{1+\left(\dot{\lambda} \Gamma \right)^2} \sigma_{z} +\left(\lambda + \frac{1}{2} \frac{d(\arctan{\dot{\lambda} \Gamma})}{dt}\right) \sigma_{x},\nonumber
\label{eq:FF}
\end{align}
where $\Gamma^{-1}=2(\lambda^2+\Delta^2)$. While this transformation is easy to construct for a qubit, it is a formidable task to find it in many-body systems~\cite{bukov2018geometric}. 

\begin{figure}[!htbp]
	\centering
	\includegraphics[width=0.5\textwidth]{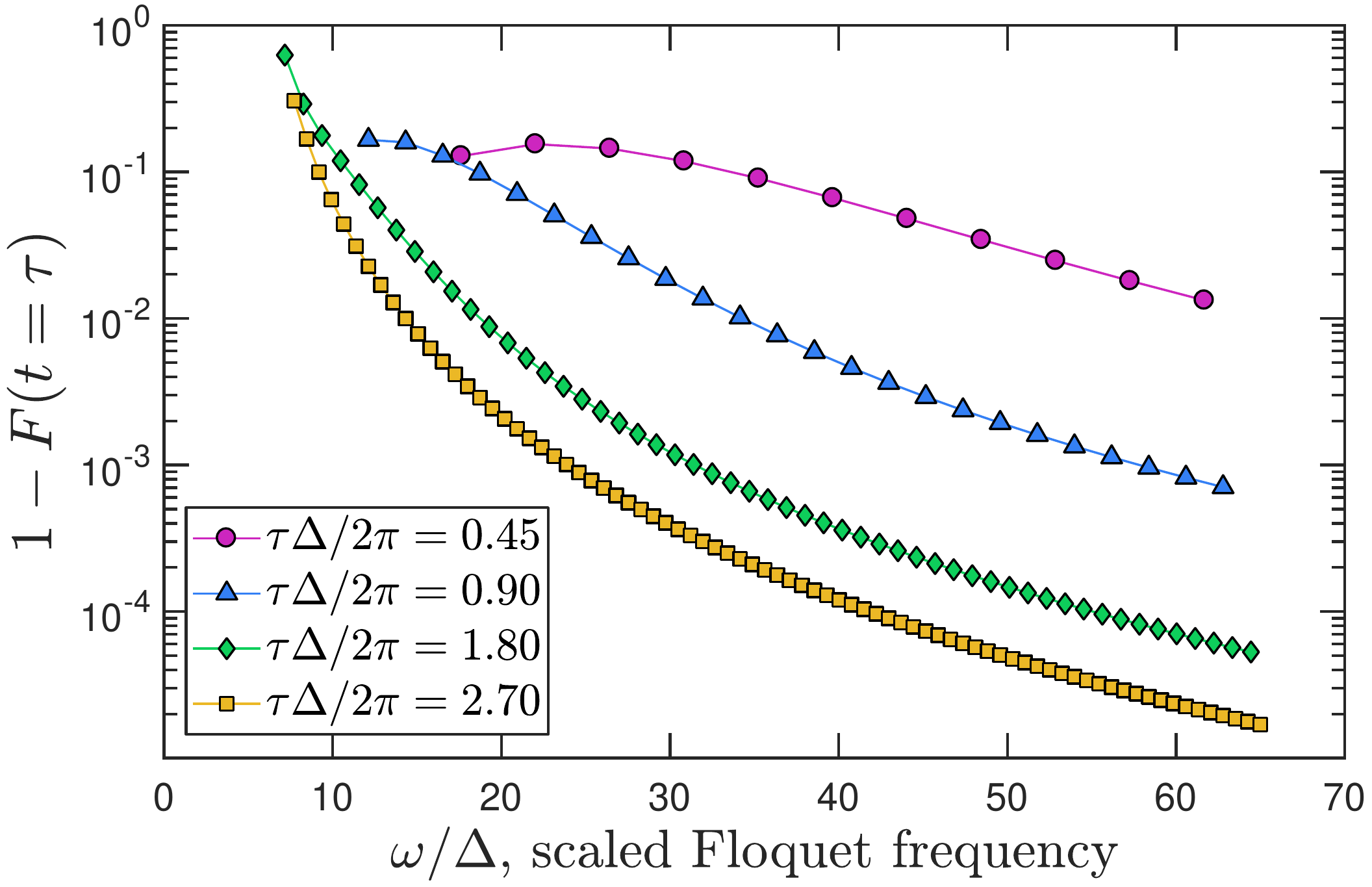}
	\caption{
		\textbf{Simulation of Floquet-engineered protocol fidelity} Numerically calculated scaling of the infidelity, $1-F$, of the Floquet-engineered protocol of duration $\tau$ with the drive frequency $\omega$ for a cubic sweep $\lambda (t) = \lambda_0(4(t/\tau)^3 - 6(t/\tau)^2 + 1)$. Points are stroboscopically sampled such that $\omega \tau = n \pi$. Protocol parameters: $\lambda_0/2\pi = 1.5\uu{MHz}, \Delta/2\pi=0.1/\mathcal{J}_0 (2 \Omega)$ MHz, $\Omega=\pi $.
	}
	\label{fig:Fig2}
\end{figure}

Our approach to constructing the fast-forward Hamiltonian exploits the idea of Floquet engineering as a way of implementing the unavailable counter-diabatic term by using control fields with both smoothly-varying and rapidly-oscillating components: 
\begin{equation}
\begin{split}
H_{\text{FE}} &=\Delta \left(1-\frac{\mathcal{J}_0(2 \Omega)}{2 \mathcal{J}_1(2 \Omega) } \dfrac{ \dot{\lambda} \cos{\omega t}}{(\Delta \mathcal{J}_0(2 \Omega))^2+ \lambda^2} \right) \sigma_z \\ 
&+  \left(\lambda(t) + \omega \Omega \sin{\omega t} \right) \sigma_x,
\end{split}
\label{eq:FEFF}
\end{equation}
where $\omega\gg\Delta$ is the drive frequency, $\mathcal{J}_0$ and $\mathcal{J}_1$ are zero and first-order Bessel functions, and $\Omega$ is a free parameter of the drive. By building in a large separation in the time-scales representing the period of the drive and the protocol time, the time-evolution of the system within each drive period
can be described with an effective Floquet Hamiltonian, which can be computed  systematically using the Magnus expansion~\cite{bukov2015universal, eckardt2017colloquium}. The desired fast forward protocol is then obtained by matching the effective Floquet Hamiltonian with the counter-diabatic Hamiltonian (see appendix \ref{append.derivation}).
For a given protocol duration, larger drive frequency results in higher protocol fidelity, as shown in fig.~\ref{fig:Fig2}.

\section{Experiment}

We experimentally implemented the Floquet-engineered fast forward protocol in a qubit formed by the electronic spin of a nitrogen-vacancy (NV) center in diamond. The spin state of the negatively-charged NV center has a long coherence time, even at room temperature, and its electronic level structure allows robust spin polarization, manipulation, and readout~\cite{Taylor2008,Childress2014}. In order to avoid hyperfine effects due to the nitrogen nuclear spin, the experiment was operated at the magnetic field corresponding to the NV excited state level anti-crossing, where optically pumping the NV center polarizes both the NV electron and nuclear spin \cite{jacques2009dynamic} (see appendix \ref{append.expt_design}). We manipulated the NV center spin by radio-frequency fields with carrier frequency $\omega_0$ near its $|m_s=0\rangle \leftrightarrow |m_s=+1\rangle$ transition, thus implementing the effective Hamiltonian in eq.~(\ref{eq:01}) in the frame rotating at this frequency. The gap $\Delta$ was controlled by detuning $\omega_0$ from the spin transition frequency, and the parameter $\lambda$ corresponds to the amplitude of the driving field, which is swept as a function of time.

\begin{figure}[ht]
	\centering
	\includegraphics[width= 0.5\textwidth]{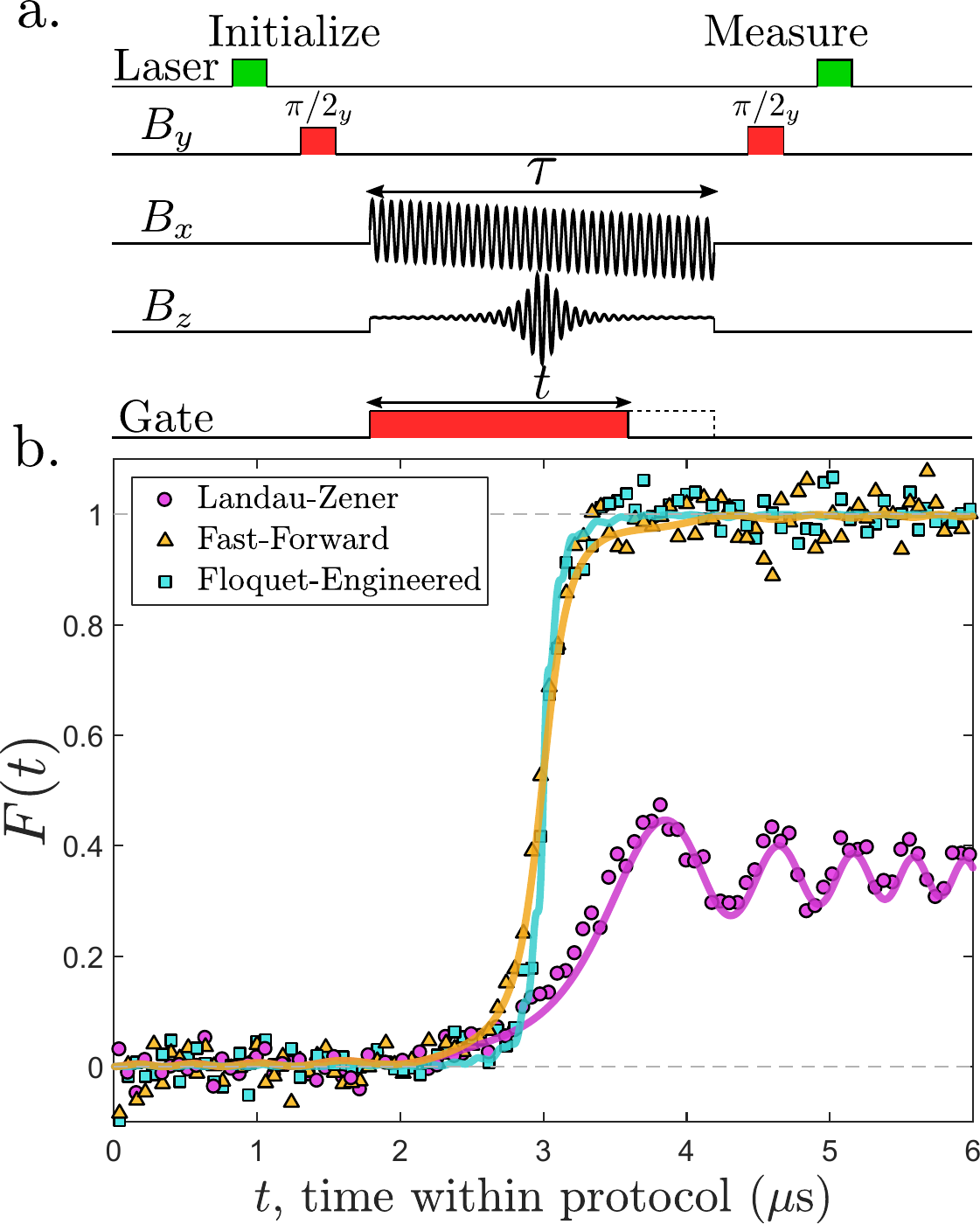}
	\caption{
		\textbf{Comparison of state preparation protocols} a. The experimental pulse sequence implementing the Floquet-engineered fast forward protocol. The qubit was manipulated using control fields $B_x,B_y,B_z$; initialization and readout was performed using laser pulses. The gate pulse was used to switch off the control fields, halting qubit state evolution after a variable protocol duration time $t$. 
		b. Measurements of fidelity during each protocol for the linear Landau-Zener sweep (purple circles), the fast-forward protocol (yellow triangles), and the Floquet-engineered protocol (blue squares). Solid lines are simulations of corresponding protocol. Oscillations in the protocols, most pronounced in Landau-Zener, are due to transitions caused by the discontinuous first derivative at the start and end of the sweep and slight misalignment of the initial state (see appendix \ref{append.expt_design}). Protocol parameters: $\lambda(t)= \lambda_0 (1-2 t/\tau)$, $\Delta/2\pi = 0.1\uu{MHz}$, $\lambda_0/2\pi = 1.5\uu{MHz}$, $\omega/2\pi =6 \uu{MHz}$, $\Omega=\pi/4$ and $\tau = 6\uu{\mu s}$.   }
		
	\label{fig:Fig2_sweeps}
\end{figure}

We performed quantum state manipulation protocols using the pulse sequence in fig.~\ref{fig:Fig2_sweeps} (a). We initialized the NV spin into the $\ket{-x}$ eigenstate with a laser pulse followed by a $\pi/2$ pulse around the y-axis. The spin then evolved under the corresponding protocol Hamiltonian, and its final spin state $|\psi\rangle$ was detected by applying another $\pi/2$ pulse around the y-axis, followed by a measurement of spin-dependent fluorescence. To track the evolution of the system throughout the protocol, we switched off the control fields after a variable time $t$, halting state evolution.

To characterize the performance of our scheme, we carried out the linear Landau-Zener sweep of $\lambda(t)/2\pi$ in the range $\pm1.5\uu{MHz}$ over time $\tau=6\mu s$ at a fixed gap $\Delta/2\pi=0.1\uu{MHz}$. The data points, together with a simulation, are shown in fig.~\ref{fig:Fig2_sweeps}. As a second benchmark, we measured the performance of the conventional fast-forward protocol by implementing the $B_z$ and $B_x$ control parameter sweeps given in eq.~(\ref{eq:FF}). The values of the gap and the linear sweep of $\lambda(t)$ were the same as for the Landau-Zener protocol. The experimentally-measured fidelity shown in fig.~\ref{fig:Fig2_sweeps} shows a dramatic improvement over the Landau-Zener protocol and approaches the value $0.990\pm0.005$ as the entire protocol is completed.

We then implemented the Floquet-engineered protocol with the time-dependence given in eq.~(\ref{eq:FEFF}), again using the same values of the parameters. Measurements of the Floquet-engineered protocol fidelity, shown in fig.~\ref{fig:Fig2_sweeps}, demonstrate that its performance closely approximates that of the conventional fast-forward protocol and its fidelity approaches $0.994\pm0.004$ as the protocol is completed.

\begin{figure}[ht]
	\centering
	\includegraphics[width= 0.5\textwidth]{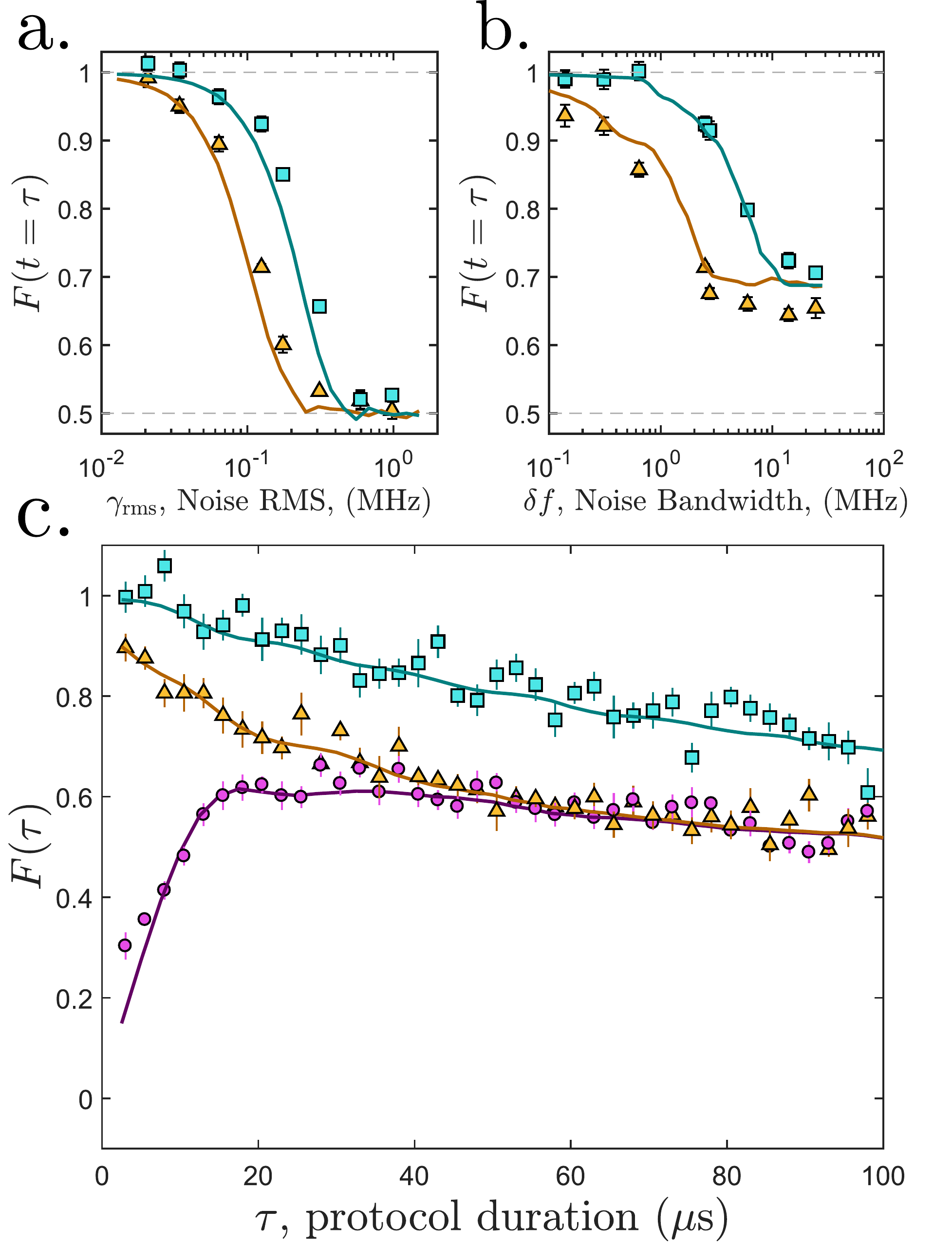}
	\caption{\textbf{State preparation in presence of noise} a. Measurements of final fidelity as a function of noise magnitude for Floquet-engineered protocol (blue squares) and conventional fast-forward protocol (yellow triangles). Noise magnitude was expressed as root-mean-square amplitude $\gamma_{rms}$ at a fixed noise bandwidth of $2.5 \, \text{MHz}$. b. Measurements of final fidelity as a function of noise bandwidth for Floquet-engineered protocol (blue squares) and conventional fast-forward protocol (yellow triangles). Noise spectral density was fixed at 0.079~MHz$/\sqrt{\text{MHz}}$. c. Final fidelity as a function of protocol duration with noise bandwidth $640\uu{kHz}$ and noise root-mean-square amplitude $64\uu{kHz}$. In all figures, solid lines are numerical simulations. Protocol parameters are the same as in fig.~3. Error bars may be obscured by data markers. }
	\label{fig:Fig4_noise}
\end{figure}

Fidelity is an important benchmark that quantifies the performance of a protocol, but to assess its potential in real-world applications it is important to study its performance in presence of the inevitable coupling to a noisy environment. We introduced such an interaction in our experiments by coupling the NV spin to a source of magnetic noise with controlled amplitude $\gamma_{rms}$ and spectral bandwidth $\delta f$ (see appendix \ref{append.expt_design}). This noise adds a stochastic term $\gamma(t)\sigma_z$ to the Hamiltonians in (\ref{eq:01}), (\ref{eq:FF}), and (\ref{eq:FEFF}), which induces transitions between the initial and final qubit states. Measurements of the fidelity of the fast-forward and the Floquet-engineered protocols in the presence of this noise are shown in fig.~\ref{fig:Fig4_noise}. The Floquet-engineered fast forward protocol is more robust to the environmental decoherence: it maintains its high-fidelity performance up to factor of 3 larger noise amplitude and factor of 5 greater noise bandwidth than the conventional fast-forward protocol with the same parameters. This can be understood by noting that the Floquet-engineered fast forward protocol performs counter-diabatic driving in the frame rotating at the Floquet frequency $\omega$. Since in this frame the noise spectrum is shifted away from zero frequency, it can efficiently induce qubit transitions within a protocol of duration $\tau$ only if it has spectral overlap with the qubit. That is, the spectral bandwidth of the noise is $\delta f \gtrsim \omega-\lambda_0/(\tau \Delta)$, where the qubit spectral bandwidth is approximately $\lambda_0/(\tau \Delta)$. This sets the noise bandwidth of approximately $5\uu{MHz}$ at which the fidelity starts to drop, as seen in fig.~\ref{fig:Fig4_noise} (b). This mechanism of protecting a qubit against environmental noise is similar to the methods of dynamical decoupling~\cite{Viola1999,DeLange2010} and further simulations demonstrating this effect can be found in Appendix \ref{append.dynamical_decoupling}. 
This argument breaks down if the noise amplitude is comparable to, or larger than, the magnitude of the $\sigma_z$ term in the corresponding Hamiltonian since the noise can no longer be treated perturbatively. 

\section{Conclusion}
Our approach demonstrates a new tool for high-fidelity quantum state manipulation in presence of environmental decoherence. The method based on Floquet engineering has the potential to be directly generalizable to high-fidelity state preparation in complex many-body quantum systems, where the counter-diabatic and the fast-forward protocols are much harder to realize. Additional promise is demonstrated by the robustness of our scheme to external noise. Our Floquet-engineering approach may find applications in a broad range of fields that rely on high-fidelity preparation of quantum states of noisy or open quantum systems, such as adiabatic quantum computing, quantum simulation, and quantum sensing beyond the standard quantum limit with entangled and squeezed states~\cite{albash218,pichler2018}.

In the late stages of our work we became aware of Ref.~\cite{petiziol2018fast} where a similar theoretical strategy of designing fast-forward protocols is proposed.

\textit{Acknowledgements:} D.S. acknowledges support from the FWO as post-doctoral fellow of the Research Foundation- Flanders. A.P was supported by NSF DMR-1813499 and AFOSR FA9550- 16- 1-0334. E.B and A.S. acknowledge support from Alfred P. Sloan foundation grant FG-2016-6728. M.P and D.K.C acknowledge support from Banco Santander Boston University-National University of Singapore grant.




\bibliography{library,ref_Floq_CD_Mohit} 

\clearpage

\appendix

\section{Derivation of FE protocol for a qubit and its generalization } \label{append.derivation}


We present here the derivation of the Floquet engineered driving protocol (equation \eqref{eq:FEFF}) for a qubit. It has the same form as that of the Fast-Forward Hamiltonian (equation~\eqref{eq:FF}):
\begin{equation}
H_{\text{FE}} =B_z (t) \sigma_z + B_x (t) \sigma_x 
\label{eq:FEFF_lab}
\end{equation}
where both $B_z$ and $B_x$ consist of smooth and rapidly oscillating parts.

Informed by the standard prescription of Floquet engineering, we take  $B_x(t) =\lambda (t)+ \omega \Omega\sin (\omega t)$, where $\Omega$ is a free parameter ~\cite{bukov2015universal}. Next, we consider a rotating frame defined by the unitary $V= \exp(- i \sigma_x \theta(t))$  where $\theta(t)=-\Omega \cos (\omega t)$, which effectively performs a re-summation of the Magnus expansion of equation (\ref{eq:FEFF_lab}). In the rotating frame, the Hamiltonian becomes:
\begin{eqnarray}
\tilde{H}&=&V^\dagger HV -i V^\dagger \dot{V} \\
&=&B_z(t) (\cos 2 \theta \sigma_z + \sin 2 \theta \sigma_y)+ \lambda (t)\sigma_x,
\end{eqnarray}
This rotating frame Hamiltonian includes $\sigma_y$, allowing us to implement the counter-diabatic (CD) Hamiltonian of equation (\ref{CD_eqn}) by choosing the appropriate time dependence for $B_z(t)$. 

To find an approximate form for $B_z(t)$, we average $\tilde{H}$ over a single time period $T= 2 \pi /\omega$ to compute the first term of its Magnus expansion. Since the $\sigma_y$ term is required for implementing CD driving, we choose $B_z(t)=\alpha- \beta(t) \cos(\omega t)$, where $\alpha, \text{and } \beta$ are free parameters of the Floquet engineered driving protocol, so that we get non-zero contribution from the average over $B_z(t)\sin 2 \theta \sigma_y$ . This gives us:

\begin{align*}
\tilde{H}^{(0)}&=  \alpha \mathcal{J}_0(2 \Omega) \sigma_z +  \beta(t) \mathcal{J}_1(2 \Omega) \sigma_y    + \lambda (t)\sigma_x 
\end{align*}
\noindent
where $\mathcal{J}_0$ and $\mathcal{J}_1$ are zero and first-order Bessel functions, and we have assumed that $\tau \gg T$ so that $\beta(t), \lambda(t)$ are approximately constant over a single period of the drive, $T$. This Hamiltonian is exactly the CD Hamiltonian~\eqref{CD_eqn} as long as the coefficients satisfy the constraint:
\begin{eqnarray}
\beta \mathcal{J}_1(2 \Omega) = \dfrac{1}{2} \dfrac{\alpha \mathcal{J}_0(2 \Omega) \dot{\lambda}}{(\alpha \mathcal{J}_0(2 \Omega))^2+ \lambda^2},
\end{eqnarray}
\noindent 
where the gap of the effective qubit in the rotating frame is $\Delta'=\alpha \mathcal{J}_0(2 \Omega)$. Note that the latter is completely arbitrary and we get a CD Hamiltonian irrespective of the value of this gap. Transforming back to the lab frame, and choosing $\alpha=\Delta$ such that the gap in lab frame remains unchanged, we arrive at our Floquet engineered driving protocol (equation \eqref{eq:FEFF}).

Additionally, to ensure that wavefunctions in both lab and rotating frames at the initial time $t=0$ and final time $t=\tau$ are identical up to a constant phase, we require that $\Omega = m \pi$ and $\tau = n T = n/{2 \pi \omega}$ for integers $m$ and $n$. If the ground state of the Hamiltonian $H_{\text{FE}}$ is in the x-direction at the initial and final times, then we have more freedom in our choice of $\Omega$ because the unitary $V= \exp(- i \sigma_x \theta(t))$ can only add an overall phase to the wavefunction (see appendix \ref{append.protocol_imperfection}). We exploited this in experiment to drive our fields at high frequency $\omega$ by taking a smaller $\Omega$, which helped us by reducing the amplitude of the required fields below the saturation level of our hardware.

The previous discussion can directly be generalized to many body systems. First we note that an optimal variational single-spin counter-diabatic protocol, which can be easily computed~\cite{SelsE3909}, can already be very efficient even in complex interacting systems~\cite{lechner} . Such variational protocols can be implemented through the Floquet fast forward driving proposed here. In a more general situation one can extend our strategy in several different ways. Here we present the most direct one. Consider a generic system subject to a time-dependent Hamiltonian:
\begin{equation}
H(t)=H_0+\lambda(t)H_1,
\end{equation}
then a Floquet engineered fast-forward Hamiltonian has the general form 
\begin{equation}
    H_{\text{FE}}=\Omega \omega \sin \omega t H+g(t)H_1.
\end{equation}
In the rotating frame defined by the unitary $V=\exp{\left(-i\Omega H \cos \omega t \right)}$, the Hamiltonian becomes 
\begin{equation}
\tilde{H}=V^\dagger H_{\text{FE}} V -iV^\dagger \dot{V}=g(t) V^\dagger H_1 V.
\end{equation}
Consequently, the first term in the Magnus expansion reads:
\begin{equation}
\tilde{H}^{(0)}=\sum_l i^l g_l \sum_{n,m} \mathcal{J}_l(\Omega (\epsilon_n-\epsilon_m)) \left< n \right| H_1 \left| m \right> \left|n \right> \left< m \right|, 
\label{eq:H_general}
\end{equation}
where $\epsilon_n$ is the eigenvalue of $H$ associated with eigenvector $\left| n \right>$, $g_l$ is the $l$th component of the Fourier series of $g(t)$ and $\mathcal{J}_l$ is the $l$th order Bessel function of the first kind. The system remains adiabatic in the rotating frame as long as $H^{(0)}$ implements the adiabatic gauge potential, given by \cite{kolodrubetz2017geometry}:
\begin{equation}
A_\lambda = \sum_{n,m} i\frac{\left< n \right| H_1 \left| m \right>}{\epsilon_m -\epsilon_n} \left|n \right> \left< m \right|.
\label{eq:AGP}
\end{equation}
In practice the number of Fourier components in $g(t)$ will be limited and the best approximation of \eqref{eq:H_general} to the adiabatic gauge potential \eqref{eq:AGP} can be found by considering $g_l$ and $\Omega$ as variational parameters and using the idea developed in~\cite{SelsE3909}. Further details on the implementation and performance of these protocols will be presented elsewhere. 


\section{Experimental design}\label{append.expt_design}

\begin{figure}[ht]
	\centering
\includegraphics[scale=0.5]{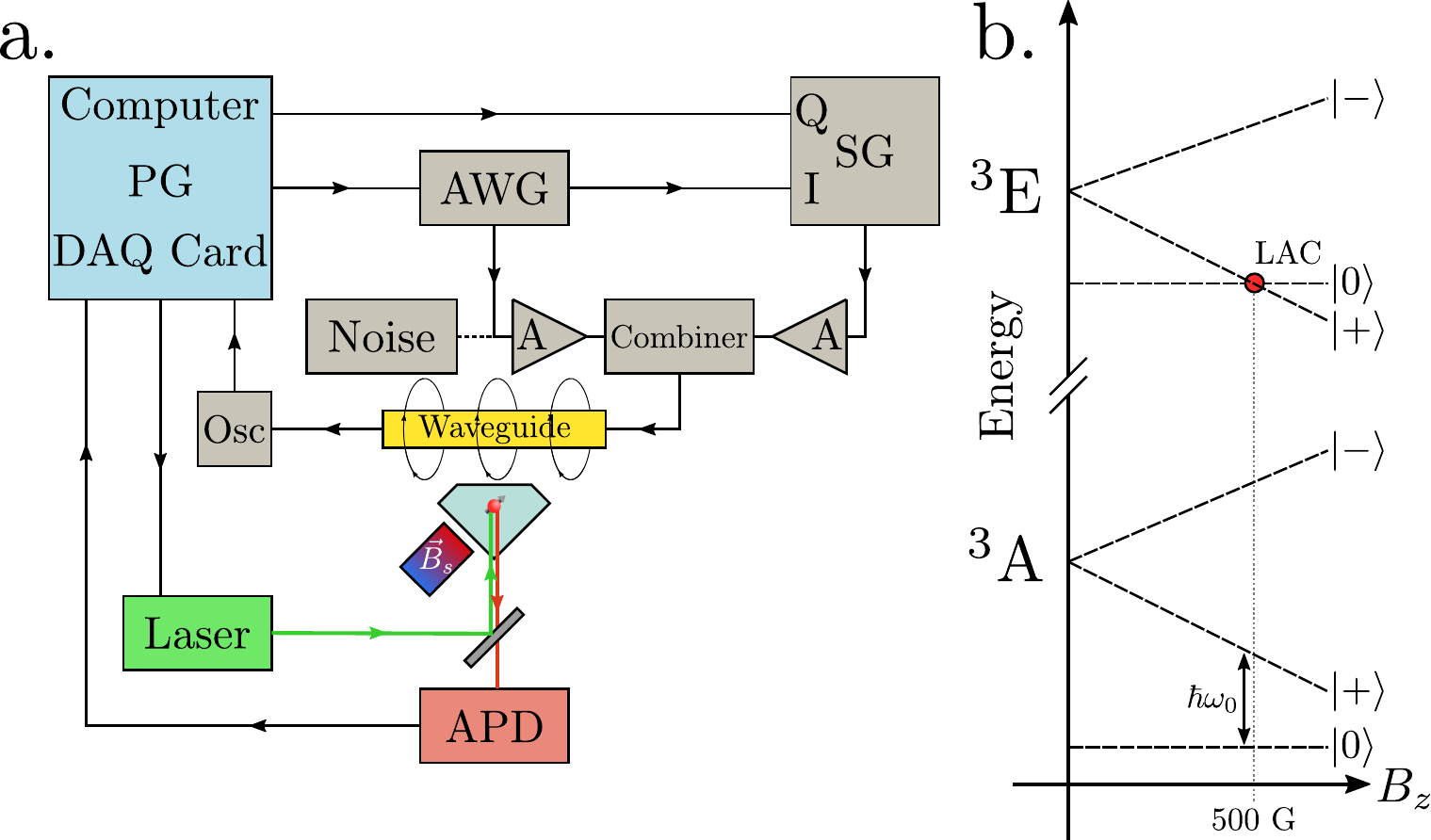}

	\caption{\textbf{Hardware and NV energy levels} a. Schematic diagram of hardware setup. PG: Pulse Generator; DAQ Card: Data Acquisition card; AWG: Arbitrary Waveform Generator; SG: Signal Generator; A: Amplifier; Osc: Oscilloscope; APD: Avalanche Photodiode. Laser module includes a double pass acoustic-optic modulator (AOM).
	b. Energy levels of the NV center under a static magnetic field along the NV symmetry axis, taken to be the z-axis.}
	\label{fig:supp_hardware}
\end{figure}

The diamond used in our experiments was grown by $\text{C}^{12}$ enriched carbon vapor deposition and bombarded with $\text{N}^{15}$ ions to produce spin-1 NV centers coupled to spin-1/2 $\text{N}^{15}$ nuclei by the hyperfine interaction $A \vec{S} \cdot \vec{I}$. Figure \ref{fig:supp_hardware}.a shows a schematic diagram of the hardware setup used to probe and manipulate individual NV centers. The setup is controlled by a computer which communicates with the hardware and has a pulse generator card (PG) for creating TTL trigger pulses and a data acquisition card (DAQ card) for receiving photon detection events from the avalanche photodiode (APD). We probe individual NV centers using a 532nm laser in a scanning confocal microscope setup using the APD to detect fluorescence and with an acoustic-optic modulator (AOM) to create laser pulses of 100ns and longer. A bar magnet ($\vec{B}_s$) mounted on a 5 axis translation/rotation stage is aligned with the NV center axis and the distance from the NV center is tuned to produce the desired static field along the NV center z-axis.

To create an effective qubit, we tuned the static field $\vec{B}_s$ to the NV center excited state level anti-crossing (LAC) at approximately 500G, as shown in the energy level diagram in Figure \ref{fig:supp_hardware}.b. At the LAC, optically pumping the NV center will polarize both the NV spin and the nuclear spin \cite{jacques2009dynamic}. Since the nuclear spin has a much longer relaxation time than the electronic spin and we do not drive at the nuclear spin transition frequency, the nuclear spin remains in the ground state throughout the protocol and the hyperfine term becomes $-A/2S_z$, merely shifting the NV electronic spin transition frequency. Additionally, at the LAC, the NV spin states $\ket{+1}$ and $\ket{-1}$ are split by ${\sim}3$ GHz, allowing us to drive on resonance with the $\ket{0} \leftrightarrow \ket{+1}$ transition at $\omega_0 \approx 1.46$ GHz without driving any transitions to the $\ket{-1}$ state. Thus, we have an effective qubit consisting of the NV spin states $\ket{0}$ and $\ket{+1}$.

\begin{figure}[ht]
	\centering
\includegraphics[scale=0.5]{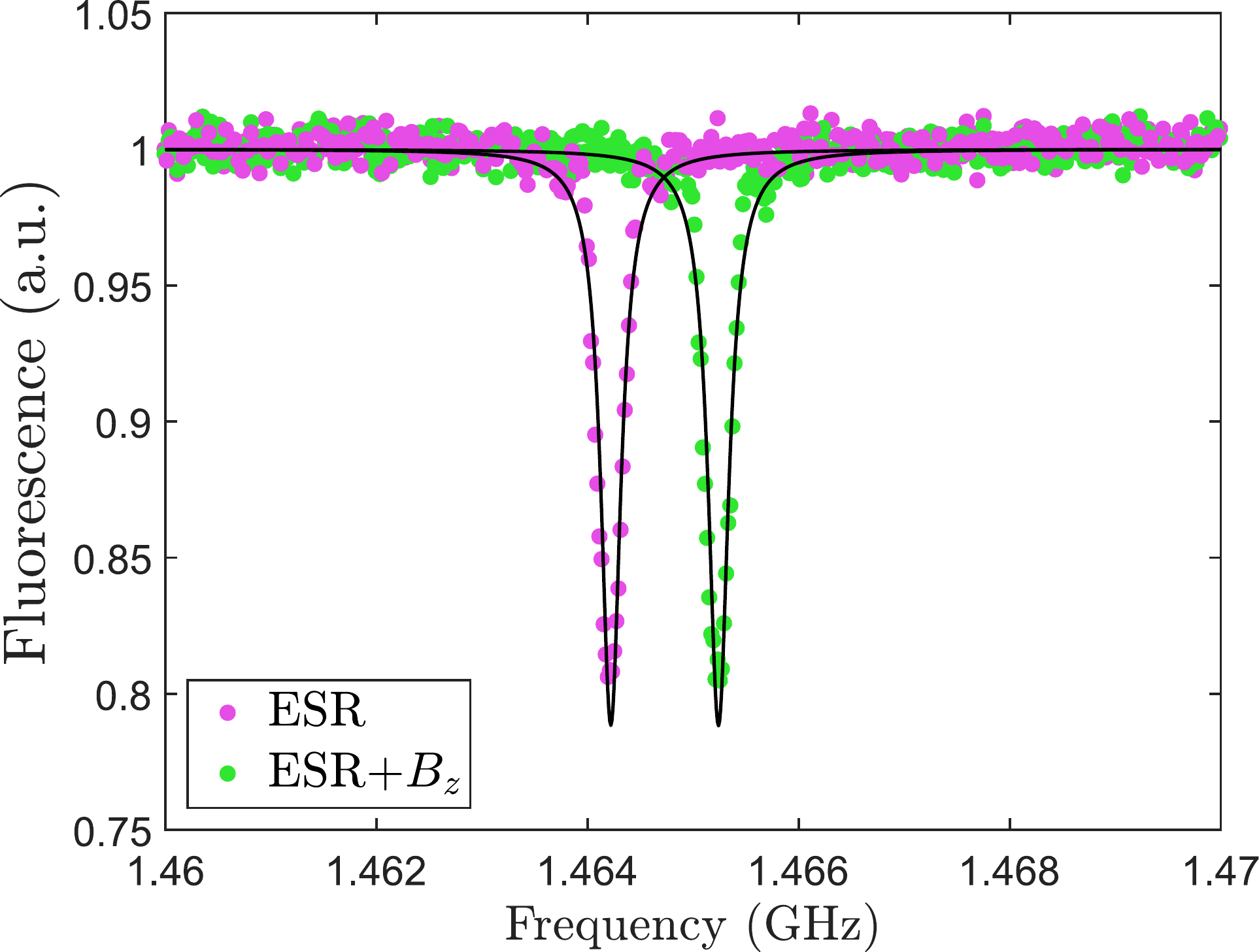}
	\caption{\textbf{ESR spectrum} Measured ESR spectrum of the NV center at the LAC with an applied $B_z = 1$MHz (green data) and without $B_z$ (purple data). Solid lines are fits to a Lorentzian line shape, $A(\omega) = 1- \frac{c}{(\omega - \omega_0)^2 + (b/2)^2}$}
	\label{fig:supp_ESR}
\end{figure}

We manipulate the qubit using time-dependent external magnetic fields $B_{x,y,z}$ generated by current in a waveguide near the NV center. To generate $B_x(t)$ and $B_y(t)$, we generate voltage signals using an arbitrary waveform generator (AWG) and use them to perform I/Q modulation of a carrier signal at frequency $\omega_0$ created by the signal generator (SG). $B_z(t)$ is also generated by an AWG, but is not modulated, and the signals are then amplified, combined, and sent to a waveguide where they generate a magnetic field at the NV center. The magnetic field generated by each of these signals has components along both the x- and z-axes of the NV center, but $B_z(t)$ has frequency components up to only $\sim 100 \, \text{MHz} \ll \omega_0$, so it cannot drive transitions and has negligible effect on the x- and y-axes. Conversely, since $B_x(t)$ and $B_y(t)$ are modulated at $\omega_0$, much faster than any other scale in the system, in the rotating frame the z-axis field they contribute rapidly averages to zero, giving the experimentally accessible Hamiltonian for the effective qubit:
\begin{equation}
H_{\text{lab}} = (\omega_0/2 + B_z) \sigma_z + 2(B_x\cos{\omega_0t} + B_y\sin{\omega_0t}) \sigma_x
\end{equation}
Since the drive amplitude and detuning are much smaller than the carrier frequency, we transform to the rotating frame defined by $\omega_0/2 \sigma_z$ and invoke the rotating wave approximation to give the following Hamiltonian:
\begin{equation}
H_{\text{rot}} = B_z(t) \sigma_z + B_x(t) \sigma_x + B_y(t) \sigma_y
\end{equation}
This allows us to implement each protocol by choosing $B_{x,y,z}$ appropriately.

To calibrate the amplitudes of $B_{x,y}$, we set them to be constants to drive Rabi oscillations and tune the power to give the desired Rabi frequency. To calibrate $B_{z}$, we set it to be constant and perform electron spin resonance (ESR) to observe the shift in the transition frequency. In Figure \ref{fig:supp_ESR} we show electron spin resonance (ESR) spectra of the NV center with the static magnet $\vec{B}_s$ tuned near the LAC (purple data). In another set of data (green), we additionally apply a constant field $B_z = 1$MHz using the electronics described above for the duration of the ESR RF pulse ($4\mu$s). The ESR spectra confirm that the $\text{N}^{15}$ nuclear spin is polarized because each spectra has a single peak while the spectrum for an unpolarized spin would have two peaks separated by the hyperfine interaction $A \approx 3$MHz. Additionally, we see that applying $B_z$ results in an effective $\sigma_z$ term, shifting the ESR frequency.

The experiment is then carried out by the pulse sequence in figure \ref{fig:Fig2_sweeps}.a, as described in the main text. When reporting the final fidelity for each protocol, $F(\tau)$, we average the fidelity over the final 40ns of the protocol in order to account for jitter in signal generation. 

\section{Protocol imperfections} \label{append.protocol_imperfection}
In this section we discuss potential errors in the Floquet-engineered protocol that might arise from simplifying approximations and limitations in the hardware. We show that these errors do not significantly affect the experimentally achievable protocol fidelities. 

\begin{figure}[ht]
	\centering
\includegraphics[scale=0.3]{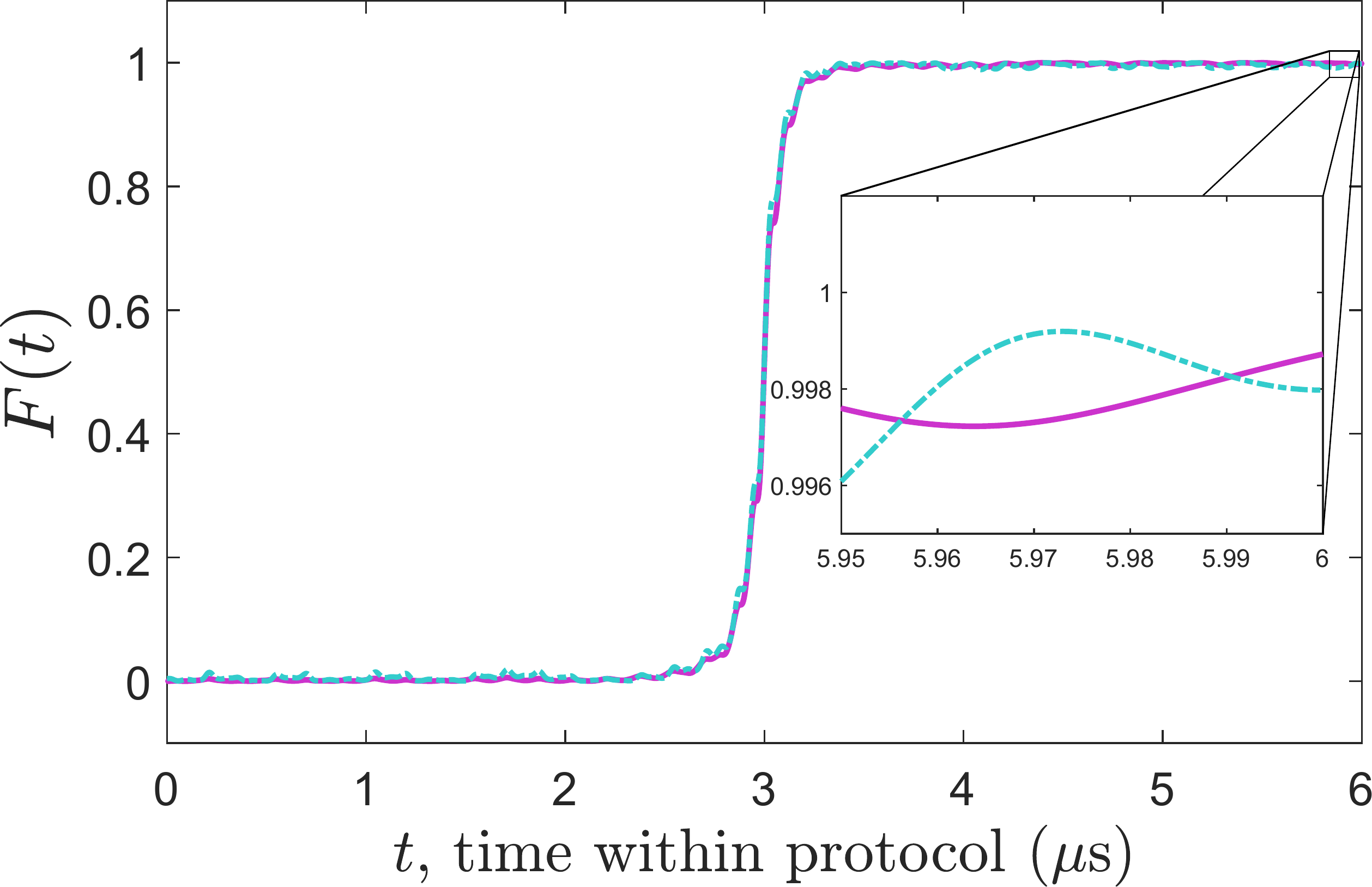}
	\caption{\textbf{ Simulation comparing fidelity for different initial states} Numerically calculate fidelity within the FE protocol with the state starting along the x-axis (solid purple line) and along the true ground state of the original LZ Hamiltonian (dashed blue line). Note that on this scale the curves nearly lie on top of each other. Protocol parameters: $\Delta/2\pi = 0.1\uu{MHz}$, $\lambda_0/2\pi = 1.5\uu{MHz}$, $\omega/2\pi =6 \uu{MHz}$, $\Omega=\pi/4$ and $\tau = 6\uu{\mu s}$ }
	\label{fig:supp_angle}
\end{figure}

As illustrated in the main text, the aim of each protocol is to bring the system from the initial state, $\ket{-x}$ to the target state, $\ket{+x}$. The Landau-Zener protocol achieves this by sweeping $B_x$ with a constant $B_z$ so that the ground state rotates with the net magnetic field around the y-axis from -x to +x. Since $B_z$ is finite, this would require $B_x \rightarrow \infty$, which is experimentally inaccessible. To approximate this, we consider protocols where $B_z/B_x \ll 1$ so that the spin pointing along -x is nearly in the initial ground state. Our experiments use $B_z/B_x = 0.1/1.5$, which gives an initial overlap of $| \braket{\psi(0)|\psi_{\text{GS}}(0)} | = 0.9978$. In figure \ref{fig:supp_angle}, we show simulations of the FE protocol with the same parameters as in Figure 3.b of the main text with the system starting in the exact ground state and along the -x-axis. The curves nearly overlap for the entire protocol and the final fidelities agree at the level of precision available in experiments $(\pm 0.004)$. The oscillations in both protocols are caused by the finite Floquet driving frequency and deviations from starting in the initial ground state; these small fluctuations will be slightly different for the different initial states.

\begin{figure}[ht]
	\centering
	\includegraphics[scale=0.35]{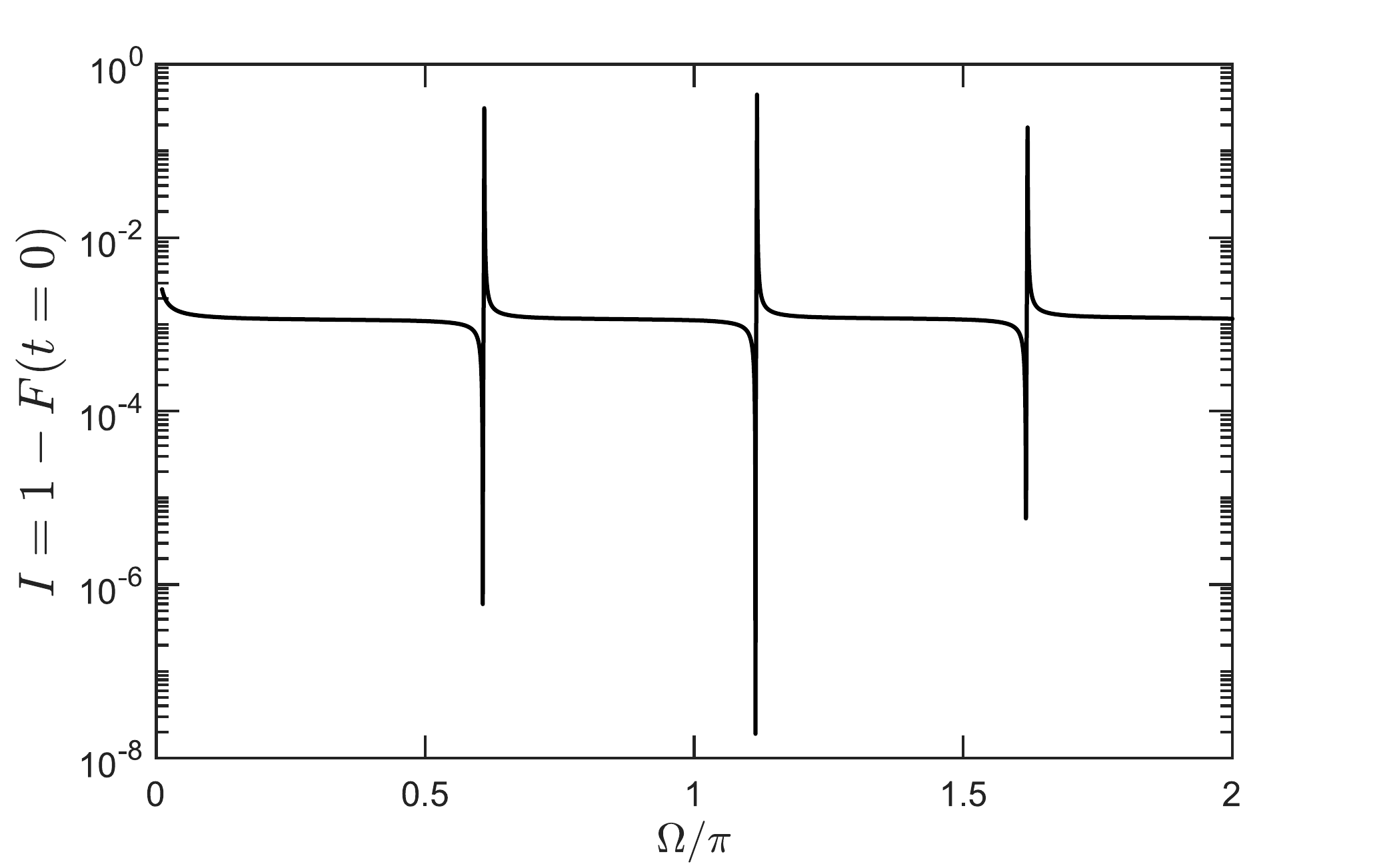}
	\caption{\textbf{Infidelity as a function of $\Omega$} Numerically calculated infidelity ($1-F$) of the initial ground state of the FE Hamiltonian with the initial state, $\ket{+x}$, as a function of the parameter $\Omega$.}
	\label{fig:supp_Omega}
\end{figure}

Another point mentioned in the main text is the choice of the parameter $\Omega$ in the FE Hamiltonian. This parameter appears in the rotating frame transformation operator $V = \exp{(i \sigma_x \Omega \cos{\omega t})}$, and hence in the lab frame FE Hamiltonian. As mentioned in the appendix \ref{append.derivation}, for the initial states in the lab and rotating frames to agree at time $t=0$, we require that $\Omega = n \pi$, for an integer $n$. However, if the initial ground state is along the x-axis, this operator does not rotate the state and merely adds an overall phase, meaning there is no restriction on $\Omega$. Freedom in choosing $\Omega$ is useful from an experimental point of view because the Floquet driving term, $\omega \Omega \sin{\omega t}$, is easier to implement if the amplitude can be made smaller by taking a smaller value of $\Omega$. As explained above, the initial ground state is not exactly along the x-axis, but instead slightly above it in the x-z plane. Thus, taking $\Omega \neq n\pi$ could result in errors in the initial state being different from the initial ground state. 

\begin{figure*}[ht]
	\centering
	\makebox[\textwidth][c]{\includegraphics[width= 0.8\textwidth]{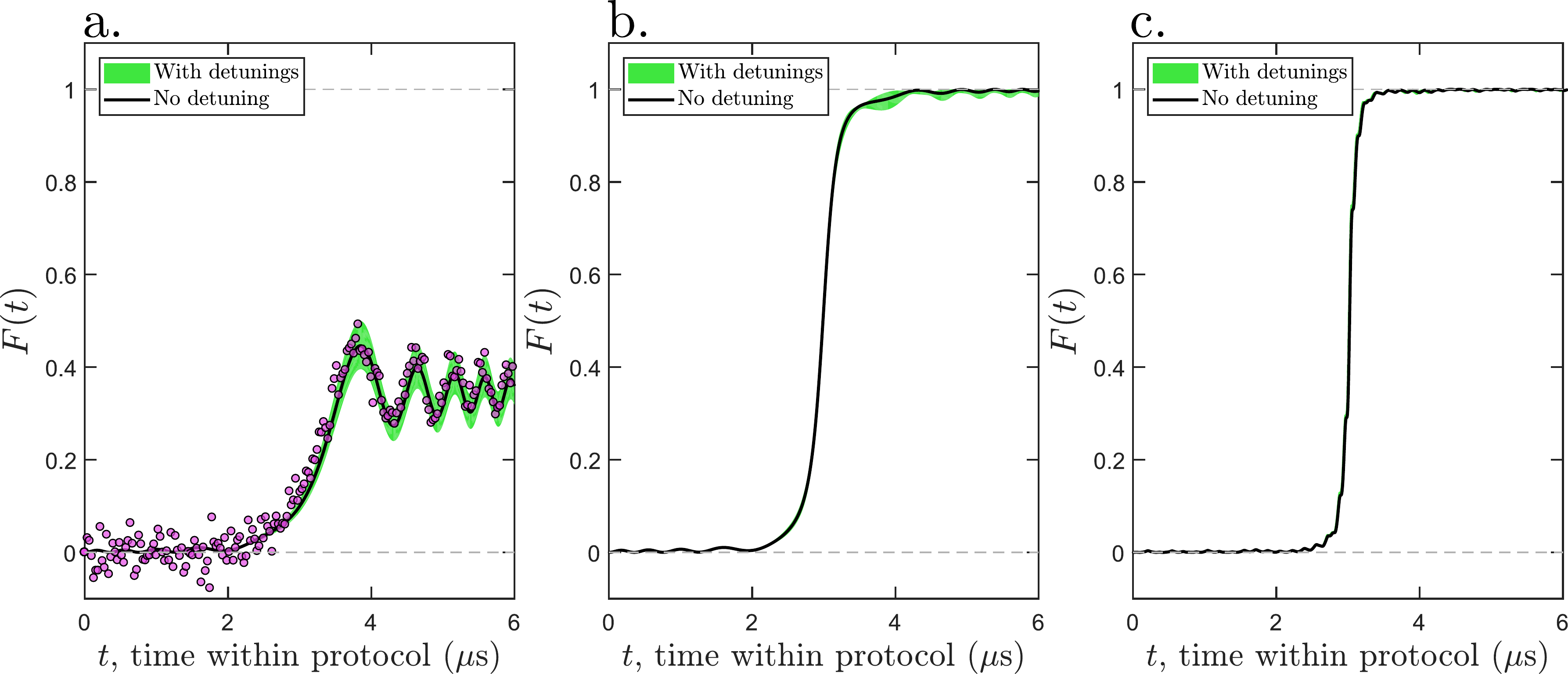}}
	\caption{\textbf{Protocol performance with detuning} a. Simulation of Landau-Zener protocol with detunings $\delta \sigma_z$ drawn from a Gaussian distribution $\delta \sim N(\mu = 0, \sigma = 8kHz)$. The simulation is repeated many times and the results averaged together with the green band capturing the mean fidelity $\pm$ 1 standard deviation at each point in time. Black curve is the simulation with no detuning. Data is plotted as well to show it falls within the green band. Plots b. and c. are the same as a. for the FF and FE protocols, respectively. Note that the uncertainty band for the FE protocol is not visible at this scale. Data points are omitted since the spread in the data is larger than the bands. Parameters are the same as for Figure \ref{fig:supp_angle}.} 
	\label{fig:supp_zOffset}
\end{figure*}

To show that we can still choose $\Omega$ freely without introducing significant infidelities, we computed the infidelity of the initial state, $\ket{-x}$, with the initial ground state of the Floquet-engineered Hamiltonian as a function of $\Omega$, shown in Figure \ref{fig:supp_Omega}. We see that for almost all values of $\Omega$, the states have infidelities of $I = 1- |\braket{-x|\psi_{\text{GS}}(t=0, \Omega)}|^2 \leq 0.002$, consistent with the infidelities we expect based only on the fact that $\Delta/\lambda_0 \neq 0$, as discussed above. The isolated points where the infidelity changes rapidly occurs when $\Omega$ approaches points such that $J_1(2 \Omega) = 0$. Near these points, we can approximate $B_z \sim \Delta - c/J_1(2 \Omega)$, for a constant $c \ll \Delta$. As $\Omega$ increases from below, it approaches $\Delta \approx c/J_1(2 \Omega)$, so that $B_z = 0$ and the ground state points along the x-axis, matching the initial state. Increasing $\Omega$ slightly more results in a large $B_z$ and the ground state points nearly along the -z-axis, giving large infidelities. As long we avoid these points when choosing $\Omega$, the initial infidelity will be small and we can expect the final infidelity to be comparably small, as in Figure \ref{fig:supp_angle} where we examined the effect of small infidelities from the initial state. 

A final imperfection we consider is detuning from the transition frequency $\omega_0$. As shown above, we perform electron spin resonance (ESR) to determine $\omega_0$ and then set the signal generator to this frequency. If the applied fields $B_{x,y}$ are detuned from resonance by a small amount $\delta$, then in the rotating frame there is an additional term $\delta \sigma_z$. In our experiment, detunings result from two main sources. First, changes in the temperature or humidity of the laboratory cause drift in the distance of the NV center from the static magnet $B_s$, shifting the ESR frequency. We observe drift in $\omega_0$ of no more than 50-100kHz on the timescale of a day. To avoid detuning resulting from this drift, we perform ESR measurements at regular intervals of 20-60 minutes during experiments and retune the frequency.

The second source of detuning is the uncertainty in the measured frequency $\omega_0$. After performing ESR, we fit the normalized data to Lorentzian lineshapes, which results in some numerical uncertainty in the fit parameters. Defining the uncertainty as half the width of the 66\% confidence interval of the fit parameters, we find typical uncertainties in $\omega_0$ of $\pm 6$-$10$kHz. We investigate how this might affect each proctol by running a simulation with an additional term $\delta \sigma_z$ where $\delta$ is drawn from a Gaussian distribution of mean zero and standard deviation 8kHz, and then averaging together the fidelities as a function of time for many iterations with independent values of $\delta$. The results in Figure \ref{fig:supp_zOffset} show that the Landau-Zener protocol is most sensitive as the uncertainty band is largest and that detunings might explain some of the deviation of the data from the simulation without detunings. We also see that the FE protocol is more robust against detunings than the FF protocol and is unaffected at this scale.

\section{Noise}
\subsection{Experimental details}
To generate classical magnetic field noise, we applied several amplification stages to the Johnson noise of a resistor at room temperature to produce white noise band-limited by the amplifiers. Because the bandwidth of the amplifiers was 300MHz, much less than the transition frequency $\omega_0$, the noise cannot drive $|0\rangle \leftrightarrow |+1\rangle$ transitions and is well described by pure dephasing: $H_{\text{noise}} = \gamma(t) \sigma_z$. The noise signal was combined with $B_z(t)$ to deliver it to the waveguide where it creates a magnetic field.

We characterized the noise by its amplitude spectral density, which we varied by adding attenuators, and its bandwidth, which we varied by adding low pass filters. We used commercially available 5th order elliptic filters which have fast rolloff of >20dB/octave, allowing us to approximate them as ideal low pass filters with constant spectral density and a hard cutoff which we define as the -3dB point. We measure the RMS of the noise generator using a digital oscilloscope and then apply the calibration to determine the magnitude of $B_z$ as explained above.

\begin{figure}[ht]
	\centering
	\includegraphics[scale=0.4]{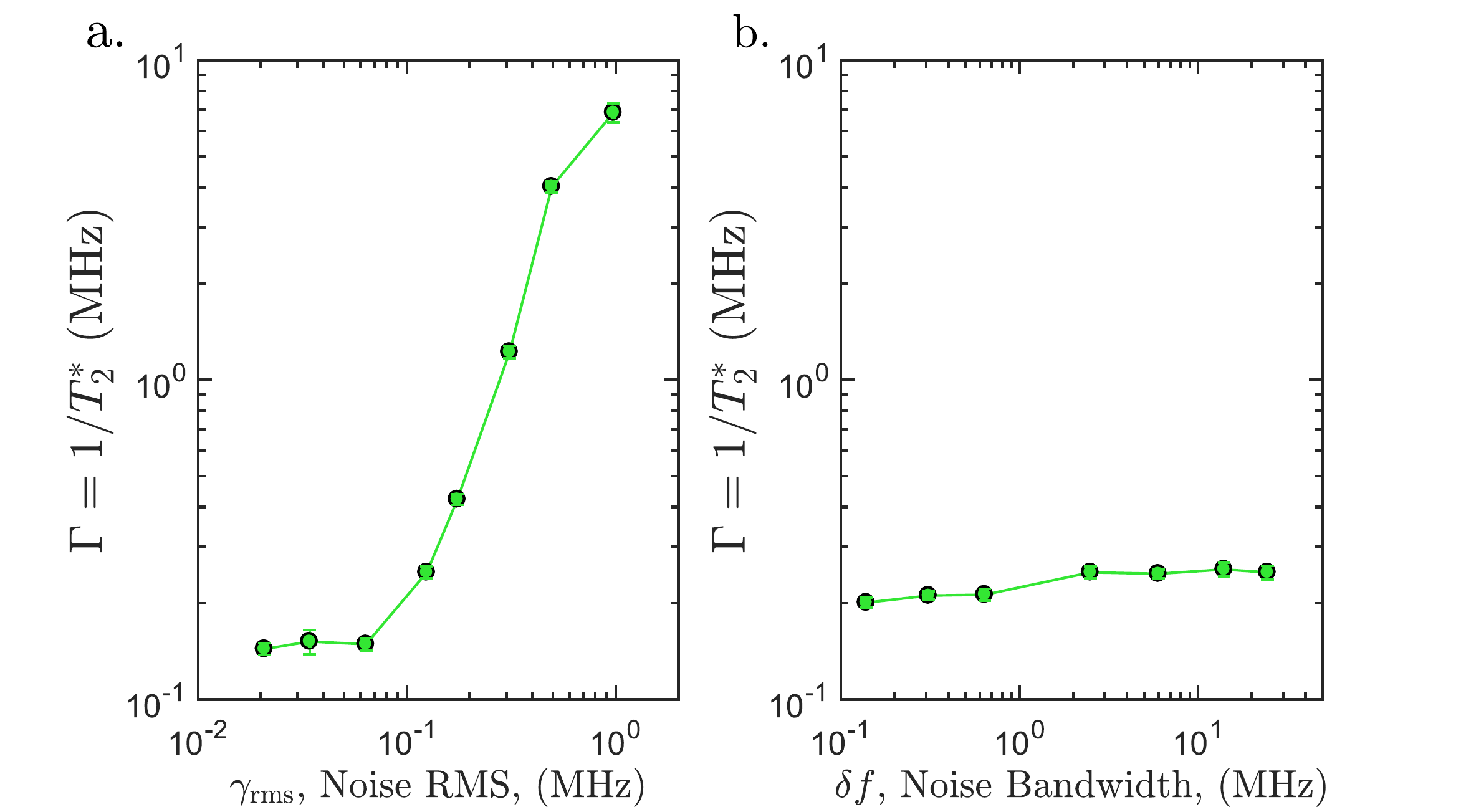}
	\caption{\textbf{Decoherence rate with external noise} a. Decoherence rate as a function of the amplitude of added noise at a fixed bandwidth of 2.5MHz. At small amplitudes, the dominant contribution to decoherence is from sources intrinsic to this NV center, and hence independent from the external noise. b. Decoherence rate as a function of the added noise bandwidth at fixed spectral density. $T_2^*$ is most sensitive to low frequency noise and is hence almost independent from the large bandwidths we consider.}
	\label{fig:supp_ramsey}
\end{figure}

To check this calibration and characterize the noise, we performed a detuned Ramsey experiment with the external noise and fit the envelope of the Ramsey fringes to an exponential decay to measure $T_2^*$. The data, reported as the dephasing rate $\Gamma = 1/T_2^*$ is reported in fig.~\ref{fig:supp_ramsey}, for the RMS and bandwidths used in fig. 4. of the main text. Fig.~\ref{fig:supp_ramsey} (a) shows that as the RMS increases, the coherence time decreases. The coherence time approaches a constant value as the added noise becomes weak and is dominated by electronic noise independent of the added noise and the dephasing resulting from the spin environment of the NV center. We find that with no added noise, $\Gamma = 0.125 \pm 0.003$ MHz, consistent with the value the data is approaching. In fig.~\ref{fig:supp_ramsey} (b), plotted on the same scale as fig.~\ref{fig:supp_ramsey} (a), we see that $T_2^*$ is nearly independent of frequency. This is expected since $T_2^*$ is most sensitive to low frequency noise, so increasingly higher frequencies get averaged out and do not affect the dynamics.

We then simulated the Ramsey experiment using the same bandwidth and adjusted the noise amplitude spectral density (ASD) until the simulated $T_2^*$ agreed with the experimental value. For a particular filter with bandwidth 2.5MHz, used in fig. 4 (a) of the main text, we find that the value of the simulated ASD required to match the $T_2^*$ is approximately 28\% larger than that estimated using the oscilloscope measurement, represented as a factor $\alpha = 1.28$. This factor can be understood as a correction to the measured RMS to account for the fact that the actual filters used do not have ideal filter shapes, meaning the model ignores noise above the -3dB point and slightly underestimates the power in the bandpass region at frequencies below the -3dB point. As a result, this correction factor will differ for different filters.

We applied this correction factor to the simulations in fig. 4 (a, c) of the main text because they use a single filter and hence are described by a single value of alpha. In fig. 4 (b), however, we use only the ASD measured by the oscilloscope since each point is taken using a different filter. As shown in fig.~\ref{fig:supp_ramsey} (b), $T_2^*$ has a weak dependence on the bandwidth at the frequencies used, so we can not reliably extract $\alpha$ for each point since $T_2^*$ gives minimal information.

\subsection{Numerical simulation details}

In the experiment, noise was characterized by spectral bandwidth $\omega_c$ and RMS. Keeping this in mind, for numerical simulations we define the noise as:

\begin{equation}
\gamma(t)=  \sqrt{ \dfrac{ 2 \omega_c \Gamma }{N}} \sum_{j=1}^{N}  \cos{ (\omega_j t + \phi_j)}
\end{equation}
where $\Gamma$ is the noise spectral density and $\omega_j$ represents different allowed frequencies within a certain bandwidth $\omega_c$. $\phi_j$ and $\omega_j$ are chosen from random uniform distribution with  $ \phi_j \in [0, 2 \pi ) $  and $\omega_j \in [0, \omega_c)$. We note that the RMS amplitude is $ \gamma_{RMS}= \sqrt{\omega_c \Gamma}$.

\section {Dynamical decoupling effect in FE driving protocol} \label{append.dynamical_decoupling}
As mentioned in the main text, the FE protocol protects the qubit from environmental noise as long as the noise spectral bandwidth is well separated from the spectral bandwidth of the driving protocol, which is simply given by its Fourier transform. In Fourier space, the noise spectrum is centered around zero with a bandwidth $\omega_c$, while the $B_z$ term of $H_{FE}$ (equation (4) in the main paper) is centered around the Floquet frequency.  Its spectrum is given by the Fourier transform of:
\begin{equation}
(\Theta(t)-\Theta(t-\tau)) \cos{\omega t} \dfrac{ \dot{\lambda} }{\Delta^2+ \lambda(t)^2} 
\end{equation}
where $\Theta(t)$ is Heaviside step function and $\omega$ is the Floquet frequency. The box function $\Theta(t)-\Theta(t-\tau)$ arises because our protocol is applied for the time $t \in [0, \tau]$. For linear ramps, we have $\lambda(t) =\lambda_0 (1-2t/\tau)$.


The total spectral function is a convolution of the Fourier transforms of the box function, cosine, and the factor $\dfrac{ \dot{\lambda} }{\Delta^2+ \lambda(t)^2}$, which for a linear ramp is a Lorentzian. Their Fourier transforms are a $sinc$ function, Dirac delta function peaked at the Floquet frequency $\omega$, and an exponential, respectively. The characteristic width of the $sinc$ scales inversely proportional to the protocol duration as $1/\tau$, while the Fourier transform of the Lorentzian decays in Fourier space over a typical scale $\frac{\lambda_0}{\tau\Delta}$. Here we are interested in the limit where $\lambda_0 \gg \Delta$, so the convolution with the $sinc$ is irrelevant and the protocol spectrum is approximately an exponential centered around $\omega$ with a characteristic decay rate $\frac{\lambda_0}{\tau\Delta}$. FE protocols are thus protected from noise as long as $\omega_c \ll \omega - \frac{\lambda_0}{\tau\Delta}$. 

\begin{figure*}[!htbp]
	\centering
	\makebox[\textwidth][c]{\includegraphics[width= 0.8\textwidth]{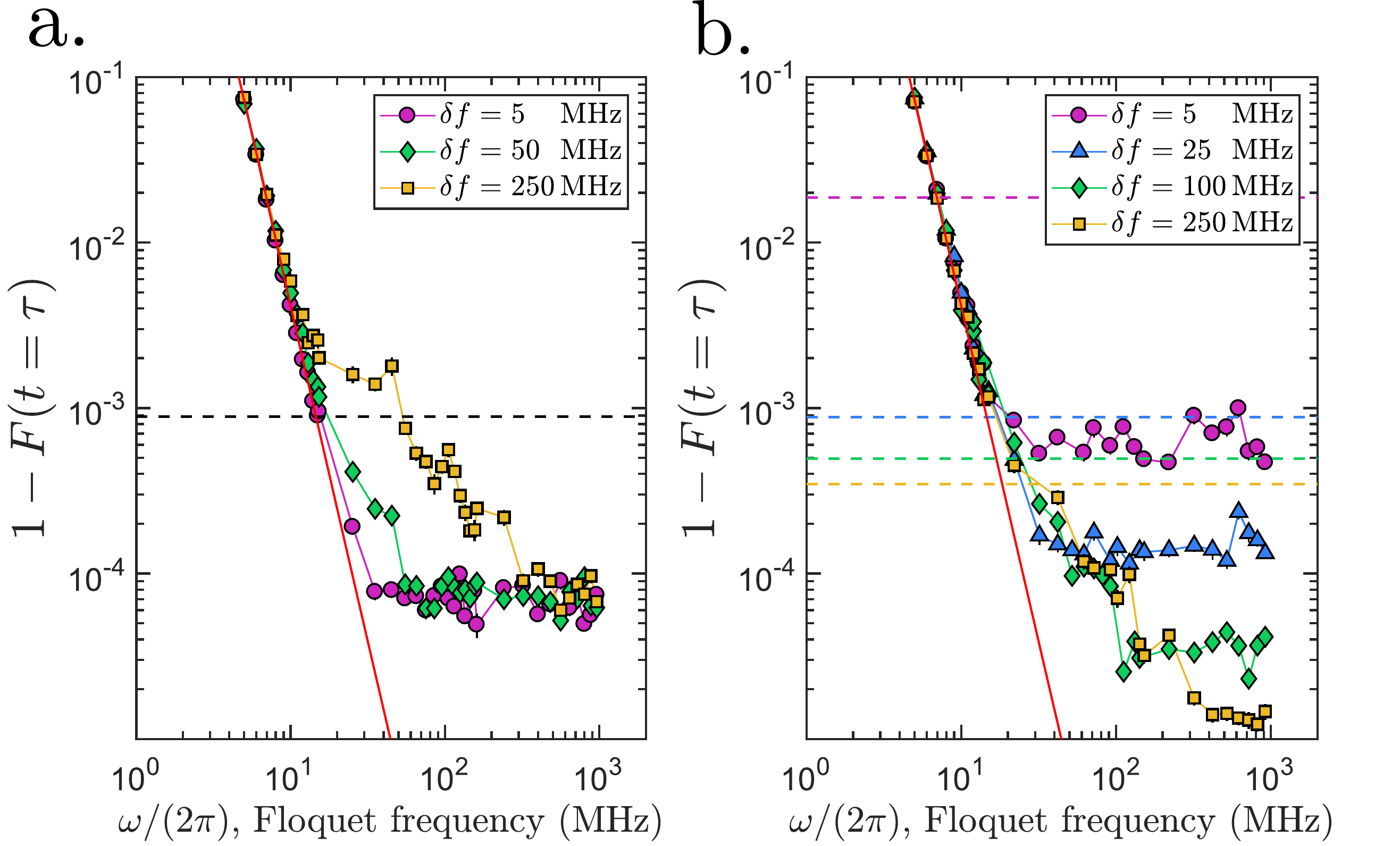}}
	\caption{\textbf{Simulations of dynamical decoupling effect} a. Numerically computed infidelity for Floquet-engineered driving with a cubic protocol as a function of stroboscopically sampled Floquet frequency, $\omega= 2 \pi n/\tau$, where $n$ is a positive integer. Each curve corresponds to a noise spectra with different bandwidths but equal spectral density. The black, horizontal dashed line is the fidelity for the conventional FF protocol, which was approximately independent of the noise bandwidth in the regime studied. The red line is the infidelity for a Floquet-engineered protocol with no noise. b. The same simulations as part a., but with constant RMS amplitude. Horizontal lines are the fidelity of the conventional FF protocol for the noise spectrum with the corresponding color.  Protocol parameters: $\lambda(t)= \lambda_0(4(t/\tau)^3 - 6(t/\tau)^2 + 1), \Delta/2\pi=0.1/\mathcal{J}_0 (2 \Omega) \uu{MHz} , \Omega=\pi$, $\lambda_0/2\pi = 1.5\uu{MHz}$ and $\tau = 4 \uu{\mu s}$ } 
	\label{fig:supp_dynamical}
\end{figure*}

To demonstrate this, we performed simulations of the Floquet-engineered protocol where we apply the noise function described above and repeat the simulation for many realizations of the noise, averaging the fidelities together. The results in fig. \ref{fig:supp_dynamical} show that for large enough Floquet frequency, the Floquet-engineered protocol (data markers and solid lines) gives lower infidelity than the conventional FF protocol (horizontal dashed lines). The infidelity decreases with increasing Floquet frequency until the Floquet frequency reaches $\omega \approx \omega_c + \lambda_0/(\tau\Delta)$, and the Floquet driving can no longer further decouple the system from the noise and the infidelity saturates.

In fig. \ref{fig:supp_dynamical} a), as we increase the noise bandwidth $\omega_c$ while keeping the spectral density $\Gamma$ constant, we find that with a large enough Floquet frequency, the Floquet-engineered protocol can give the same infidelity when the noise bandwidth is increased. This shows that, like dynamical decoupling, the Floquet-engineered protocol protects the qubit from noise as long the Floquet frequency is larger than $\omega_c$. In fig. \ref{fig:supp_dynamical} b),  as we increase the noise bandwidth $\omega_c$ while keeping $\gamma_{RMS}$ constant, we see that the infidelity of the FF protocol decreases because it is more sensitive to lower frequencies and the spectral density must decrease to give constant $\gamma_{RMS}$. However, by increasing the Floquet frequency, the Floquet-engineered protocol can achieve smaller infidelities and saturates approximately when $\omega \approx \omega_c + \lambda_0/(\tau\Delta)$, consistent with the data for constant noise spectral density. Thus, the Floquet-engineered protocol can protect the system from noise by driving at high frequency.
\end{document}